\documentclass{ws-ijmpe}
\usepackage[super,compress]{cite}
\def\nn{\nonumber}
\newcommand{\be}{\begin{equation}}
\newcommand{\ee}{\end{equation}}
\newcommand{\bea}{\begin{eqnarray}}
\newcommand{\eea}{\end{eqnarray}}

\newcommand{\om}{\omega}

\newcommand{\vk}{\vec k}

\newcommand{\vl}{\vec l}

\newcommand{\del}{\partial}

\begin{document}

\markboth{Sabyasachi Ghosh}{Thermal conductivity of hot pionic medium}

\catchline{}{}{}{}{}

\title{Thermal conductivity of hot pionic medium due to
pion self-energy for $\pi\sigma$ and $\pi\rho$ loops}

\author{Sabyasachi Ghosh}

\address{Instituto de Fisica Teorica, Universidade Estadual Paulista,\\ 
Rua Dr. Bento Teobaldo Ferraz, 271, 01140-070 Sao Paulo, SP, Brazil\\
sabyaphy@gmail.com}

\maketitle


\begin{abstract}
The thermal conductivity of pionic medium has been evaluated with the 
help of its standard expression from the relaxation time approximation,
where inverse of pion relaxation time or pion thermal width has been
obtained from the imaginary part of pion self-energy. 
In the real-time formalism of thermal field theory, the finite temperature
calculations of pion self-energy for $\pi\sigma$ and $\pi\rho$ loops have been done.
The numerical value of our thermal conductivity
increases with temperature very softly, though at particular
temperature, our estimation has to consider a large band of phenomenological uncertainty.
\end{abstract}


\ccode{PACS numbers:}

%
%
%
%
%
\section{Introduction}
\label{sec:intro}
In order to describe the expanding fire ball, produced at RHIC or LHC,
the inclusion of dissipative effects in hydrodynamical and transport
simulations has attracted remarkable attention in the recent years.
\begin{table}[pt]
\tbl{Numerical values of thermal conductivity $\kappa$ of some
earlier works in the hadronic temperature domain $T\approx0.12-0.17$ GeV
or near the transition temperature.}
{\begin{tabular}{@{}cc@{}} \toprule
Earlier works (adopted model or method) & $\kappa$ (GeV$^2$) \\
\hline
Ref.~\cite{NJL_Marty} (NJL model) & 11.6-7.78    \\
Ref.~\cite{NJL_Marty} (DPQM model) & 4.5   \\
Ref.~\cite{Nam} (Liquid-instanton model) &  0.12   \\
Ref.~\cite{Juan} (unitarization method) &  0.07-0.1   \\
Ref.~\cite{Sarkar_kap} (CEA method) &  0.032-0.049 \\
Ref.~\cite{Sarkar_kap} (RTA method) &  0.016-0.02  \\
Ref.~\cite{Greiner} (BAMPS, $\sigma\approx$43 mb) & 2.59/$\sigma~\approx$~0.023 \\
Ref.~\cite{Davesne} (CEA method) & 0.015-0.024   \\
Ref.~\cite{Gavin} (RTA method)    &  0.016-0.024  \\
Ref.~\cite{Prakash} (CEA method)   &  0.016-0.024   \\
Ref.~\cite{Nicola} (unitarization method) &  0.007-0.008    \\
\botrule
\end{tabular}}
\label{tab}
\end{table}
Only the shear viscosity of 
the expanding matter is taken as a relevant dissipative coefficient
by most of the communities, doing such kind of simulations.
To explain the elliptic flow parameter, $v_2$, extracted from data
collected at RHIC and LHC, their investigations suggest that the matter
is likely to have a very small ratio of shear 
viscosity to entropy density, $\eta/s$.
As a next attempt, the inclusion of bulk viscosity
coefficient ($\zeta$) in those simulations has been started
in a few studies but the effect of thermal conductivity ($\kappa$)
on those time evolution pictures of expanding matter is
largely ignored except the recent Refs~\cite{Kapusta,Denicol},
trying to develop a model of the thermal conductivity.
This is because the thermal conduction does not take place very well in
the system, where net baryon density is approximately zero~\cite{Gavin,Greiner,Sarkar_kap}.
Most of the hydrodynamical simulations are focused
on describing the baryon-free matter, which is expected to
be created in the central rapidity region at the RHIC and LHC.
However, thermal conductivity definitely demands its
attention to describe the expanding fire ball at forward rapidities 
or at smaller collision energies,
such as those at FAIR and in the low-energy runs at RHIC.
Probably for same reason, the microscopic calculation
of thermal conductivity for the strongly interacting matter
is not extensively explored to know its explicit temperature dependence
as done rigorously for the other dissipative quantities viz. shear and bulk
viscosities. To date, only a few studies have addressed calculation of thermal
conductivity of the strongly interacting 
matter~\cite{Gavin,Greiner,Sarkar_kap,Sarkar_kap2,Nicola,Prakash,
Davesne,Nam,Juan,Mattiello,NJL_Iwasaki,NJL_Marty,CFL1,CFL2}.
Their estimations collectively exhibit a
band of predictions as shown in the Table~\ref{tab}.

Here our interest is to calculate the thermal conductivity
of pionic medium, which may also appear in between the 
chemical and kinetic freeze-out temperatures ($T_{ch}\approx 0.17$ GeV
to $T_f\approx0.12$ GeV) at zero baryon chemical potential 
of the expanding fire ball during its late stage.
Though the pionic constituent particles do not carry any baryon number
but for the medium with finite pion chemical potential $\mu_\pi$,
where total number of pion remain constant, may have a non-zero
value of thermal conductivity~\cite{Gavin,Sarkar_kap}. 
We will make our estimation for 
$\mu_\pi=0$, which will help to compare with other relevant results.

\section{Formalism}
\label{sec:form}
Let us start from the standard expression of thermal conductivity for pionic
medium in the relaxation time approximation (RTA)~\cite{Gavin} :
\be
\kappa=\frac{\beta^2 I_\pi}{6\pi^2}\int^{\infty}_{0} 
\frac{d\vk\vk^4}{{\om_k^\pi}^2\Gamma_\pi}(\om_k^\pi-h_\pi)^2n_k(\om^\pi_k)\{1+n_k(\om^\pi_k)\}~,
\label{kap_pi}
\ee
where $n_k(\om^\pi_k) =1/\{e^{\beta\om^\pi_k}-1\}$
is the Bose-Einstein (BE) distribution function of pion with 
$\om^\pi_k = (\vk^2 +m_\pi^2)^{1/2}$, and $\Gamma_\pi(=1/\tau_\pi,~\tau_\pi$
is pion relaxation time) is the 
thermal width of $\pi$ mesons in the medium. Here $I_\pi=3$ is
pion isospin factor and $h_\pi$ is the exact heat function per particle
for the pionic medium, which can be estimated from the ideal gas expression,
\be
h_\pi=(e_\pi +p_\pi)/\rho_\pi=Ts_\pi/\rho_\pi~,
\ee
where
\be
s_\pi=I_\pi\beta\int \frac{d^3\vk}{(2\pi)^3} \left(\om^\pi_k+\frac{\vk^2}{3\om^\pi_k}\right)
n_k(\om^\pi_k)
\ee
and
\be
\rho_\pi=I_\pi\int \frac{d^3\vk}{(2\pi)^3}n_k(\om^\pi_k)
\ee
are entropy density and number density respectively of the 
pionic medium at temperature $T=1/\beta$.

\begin{figure}  
\begin{center}
\includegraphics[scale=0.52]{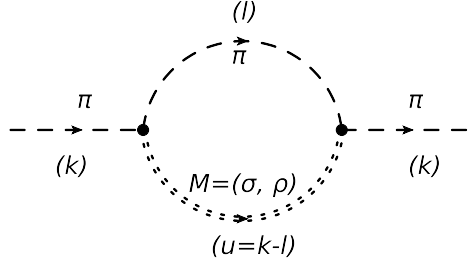}
\caption{Pion self-energy diagram for $\pi\sigma$
and $\pi\rho$ loops.} 
\label{pi_rho_loop}
\end{center}
\end{figure}

Now, the thermal width, $\Gamma_\pi$
for pion can be determined from the 
imaginary part of its one-loop self-energy~\cite{GKS,G_PRC,G_IJMPA,G_JPG}
at finite temperature, by the relation:
\bea
\Gamma_\pi(\vk,T)&=&\sum_M\Gamma_{\pi(\pi M)}(\vk,T)
\nn\\
&=&-\sum_M{\rm Im}{\Pi}^R_{\pi(\pi M)}(k_0=\om^\pi_k,\vk,T)/m_\pi~.
\label{Gam_pi}
\eea
In the above equation, ${\Pi}^R_{\pi(\pi M)}$ is the retarded component
of pion self-energy for $\pi M$ loops, where $M$ stands for $\sigma$
and $\rho$ resonances.
Feynman diagram of pion self energy for $\pi M$
loops is shown in the Fig.~(\ref{pi_rho_loop}), where $k$ is the
momentum of external $\pi$ line while $l$ and $u=k-l$ are momenta of internal
$\pi$ and $M$ lines respectively.

Following the effective hadronic model as adopted in earlier
Refs~\cite{GKS,G_JPG,SSS}, the $\sigma$ and $\rho$ resonances
in $\pi\pi$ scattering have traditionally been introduced
by using the effective Lagrangian densities for $\pi\pi\sigma$ and $\pi\pi\rho$ 
interactions: 
\be
{\cal L} = g_\rho \, {\vec \rho}_\mu \cdot {\vec \pi} \times \del^\mu {\vec \pi} 
- \frac{g_\sigma}{2} m_\sigma {\vec \pi}\cdot {\vec\pi}\,\sigma,
\label{Lag_pipiM}
\ee
where the coupling constants ($g_\sigma=5.82$ and $g_\rho=6$) are fixed from
their experimental decay widths of $\sigma$ and $\rho$ mesons
in their $\pi\pi$ channels~\cite{GKS}.
The negative sign in the second term for $\sigma\pi\pi$ coupling
is important to determine the relative phase~\cite{Harada}.
With the help of the above Lagrangian densities (\ref{Lag_pipiM}),
pion self-energy for $\pi M$ loops (shown in Fig.~\ref{pi_rho_loop}) 
with $M=\sigma$ and $\rho$ can be 
derived in the real-time formalism of thermal field theory~\cite{GKS}.
As the pion pole ($k_0=\om^\pi_k,\vk$) is situated in the Landau
cut of the self-energy function, ${\Pi}^R_{\pi(\pi M)}(k_0,\vk)$
so the corresponding thermal width can simply be expressed as~\cite{GKS,G_PRC,G_IJMPA,G_JPG}
%
\bea
\Gamma_{\pi(\pi M)}(\vk,T)&=&{\rm Im}{\Pi}^R_{\pi(\pi M)}(k_0=\om^\pi_k,\vk,T)/m_\pi
\nn\\
&=& \frac{1}{m_\pi}\left[ \int\frac{d^3{\vec l}}{(2\pi)^3}
L(l_0=-\om^\pi_l,\vl,k)\{n_l(\om^\pi_l)
\right.\nn\\
&&\left. 
- n_u(\om^M_u)\}\delta(k_0+\om^\pi_l-\om^M_u)\right]_{k_0=\om^\pi_k}~,
\label{G_pi_piM}
\eea
where $n_l$ and $n_u$ are BE distribution functions of $\pi$ and $M$
mesons respectively. 
%
%
%
Using the effective Lagrangian densities (\ref{Lag_pipiM}), one can obtain the 
vertex factors:
\bea
L(k,l) &=& - \frac{g^2_\sigma m_\sigma^2}{4}, 
~{\rm for}~M=\sigma~,
\nn\\
 &=& -\frac{g^2_\rho}{m_\rho^2} \, 
[ k^2 \left(k^2 - m^2_\rho\right) + 
l^2 \left(l^2 - m^2_\rho\right) 
- \, 2\{ (k\cdot l) \, m^2_\rho + k^2 \,l^2 \}],~{\rm for}~M=\rho~.
\nn\\
\eea
A hadronic form factor $F(\vl)=\Lambda^2/(\vl^2+\Lambda^2)$ 
with $\Lambda=1$ GeV has been multiplied
with each of the effective coupling constants, $g_\sigma$ and $g_\rho$ to consider
finite size effect of hadronic vertices.

\section{Results and Discussion}
\label{sec:num}
\begin{figure}
\begin{center}
\includegraphics[scale=0.35]{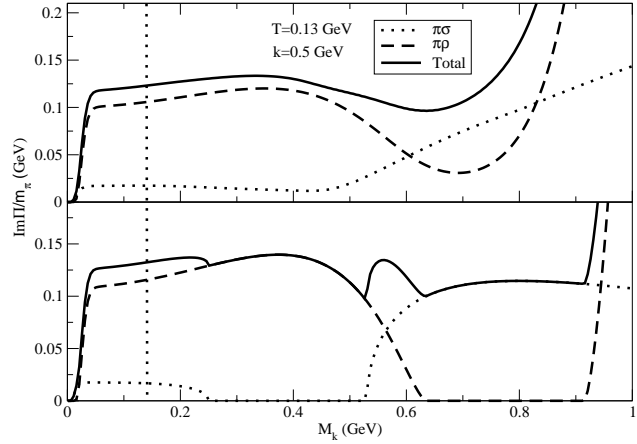}
\caption{Invariant mass distribution function, 
$\Gamma_{\pi(\pi M)}(M_k)={\rm Im}\Pi^R_{\pi(\pi M)}(M_k)/m_\pi$
for $\pi\sigma$ (dotted line), $\pi\rho$ (dashed line) loops
and their total (solid line) at fixed values of $\vk$ and $T$.
Upper and lower panels consist the results with and without 
folding effect respectively.}
\label{self_pi_M}
\end{center}
\end{figure}
\begin{figure}
\begin{center}
\includegraphics[scale=0.35]{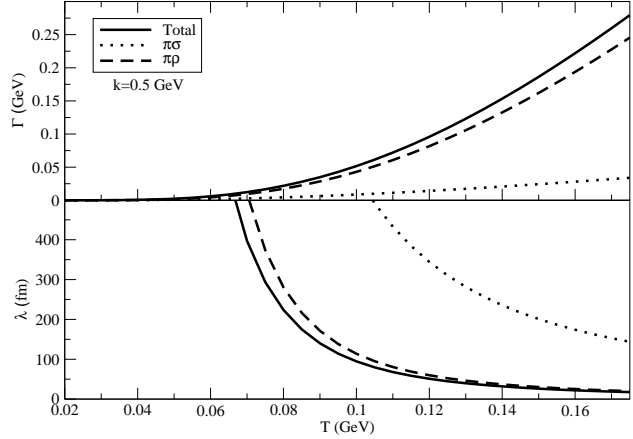}
\caption{At fixed pion momentum ($\vk=0.5$ GeV), the temperature 
dependence of $\Gamma_{\pi(\pi\sigma)}$ 
(dotted line), $\Gamma_{\pi(\pi\rho)}$ (dashed line) and their
total (solid line) are shown in the upper panel whereas lower panel demonstrates
the corresponding mean free path contributions of two individual loops and their
total.} 
\label{G_pi_T}
\end{center}
\end{figure}
\begin{figure}
\begin{center}
\includegraphics[scale=0.35]{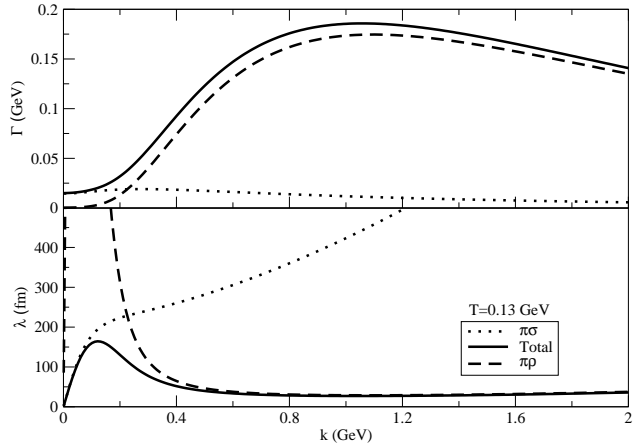}
\caption{Similar to Fig.~(\ref{G_pi_T}), for a fixed value of T against $\vk$-axis.} 
\label{G_pi_k}
\end{center}
\end{figure}
Let us start our discussion of numerical results from the pion
thermal widths, which inversely control the numerical strength
of thermal conductivity of pionic medium.
In Fig.~(\ref{self_pi_M}), the off-mass shell behavior of
$\Gamma_{\pi(\pi M)}(M_k,\vk,T)$ is plotted against the invariant
mass axis $M_k$ 
before going to analyze the on-shell function $\Gamma_{\pi(\pi M)}(M_k=m_\pi,\vk,T)$. 
The numerical values
of $\Gamma_{\pi(\pi M)}(M_k,\vk,T)$ can be generated if we 
replace $k_0=\{\vk^2+m_\pi^2\}^{1/2}$ by 
$k_0=\{\vk^2+M_k^2\}^{1/2}$ in Eq.~(\ref{G_pi_piM})
for fixed values of pion momentum ($\vk=0.5$ GeV) and temperature of the medium 
($T=0.13$ GeV).
In the lower panel of Fig.~(\ref{self_pi_M}), the dotted line is representing 
the Landau and unitary cut contributions for $\pi\sigma$ loop in the 
regions ($M_k=0$ to $m_\sigma-m_\pi=0.25$ GeV) and 
($M_k=m_\sigma+m_\pi=0.53$ GeV to $\infty$) respectively.
Similarly, the dashed line is displaying
the corresponding Landau and unitary cut contributions for $\pi\rho$ loop in the 
regions ($M_k=0$ to $m_\rho-m_\pi=0.63$ GeV) and 
($M_k=m_\rho+m_\pi=0.91$ GeV to $\infty$) respectively.
Owing to the broad spectral width of $\sigma$ and $\rho$ resonances,
we have folded the pion thermal width $\Gamma_{\pi(\pi M)}$ by the 
vacuum spectral functions of those resonances as we have done in our 
previous work~\cite{GKS} to calculate shear viscosity. 
The multi-peak
structure of total $\Gamma_{\pi(\pi M)}(M_k)$ (solid line in the lower panel) is
polished to a smooth curve (solid line in the upper panel) after 
introducing the folding effect. This is because
the region of branch cuts are generally overlapped to each other 
during the folding operation, which is reflected by the dotted and dashed lines
for $\pi\sigma$ and $\pi\rho$ loops in the upper panel of Fig.~(\ref{self_pi_M}). 
The on-shell contribution of the pion
thermal width (i.e. $\Gamma_{\pi(\pi M)}(M_k=m_\pi)$)
has been indicated by the straight dotted line in the Fig.~(\ref{self_pi_M}),
where we see that the contribution of $\pi\rho$ 
loop dominates over the $\pi\sigma$ loop and its total value becomes 
little smaller due to folding effect.

This on-shell contribution from the Landau cut is 
associated with the forward and reverse scattering processes~\cite{Weldon}. 
In the forward process, the propagating $\pi^+$ may disappear by 
absorbing a thermalized $\pi^-$ from the medium to create a 
thermalized $\rho^0$ or $\sigma$. Whereas in the reverse process, 
the $\pi^+$ may appear by absorbing 
a thermalized $\rho^0$ or $\sigma$ from the medium as well as by emitting a 
thermalized $\pi^-$.

In upper panel of Fig.~(\ref{G_pi_T}), the on-shell contribution of
$\pi\sigma$ (dotted line), $\pi\rho$ (dashed line) loops and total (solid line) 
thermal width for pion are plotted against temperature
axis. Contributions of both loops are monotonically increasing functions 
but $\pi\rho$ takes leading part in the total contribution. 
The corresponding results for mean free path, defined by 
$\lambda_{\pi(\pi M)}=\vk/(\om^\pi_k\Gamma_{\pi(\pi M)})$, are
shown in the lower panel of Fig.~(\ref{G_pi_T}).
Being inverse of thermal width, the mean free paths become decreasing 
functions of $T$. 
\begin{figure}
\begin{center}
\includegraphics[scale=0.35]{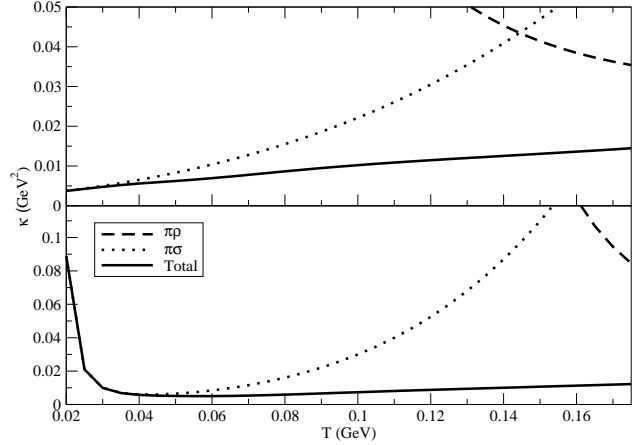}
\caption{The contributions of $\pi\sigma$ (dotted line), $\pi\rho$ (dashed line)
loops and their total in the thermal conductivity $\kappa$ are plotted against 
$T$ axis. The upper and lower panels of this figure and latter Figs.~(\ref{kap_T_comp})
and (\ref{kap_T_sigma}) are allotted for demonstrating
with and without folding results respectively.} 
\label{kap_T_srT}
\end{center}
\end{figure}
\begin{figure}
\begin{center}
\includegraphics[scale=0.35]{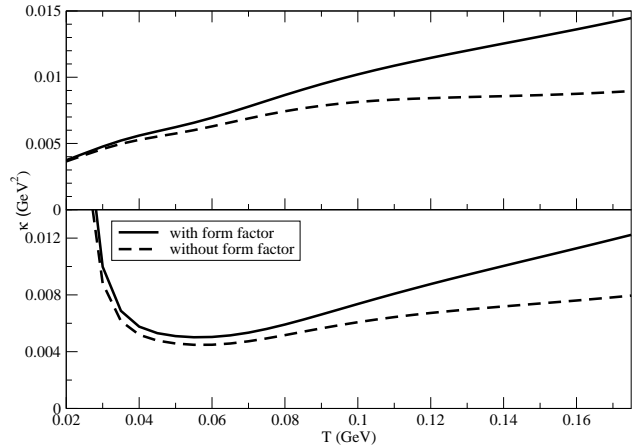}
\caption{$T$ dependence of $\kappa$ without (dashed line)
and with (solid line) form factor.} 
\label{kap_T_comp}
\end{center}
\end{figure}
\begin{figure}
\begin{center}
\includegraphics[scale=0.35]{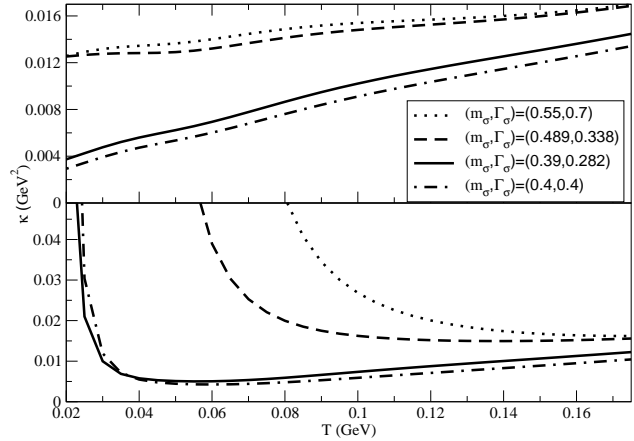}
\caption{Total $\kappa(T)$ for different set of mass $m_\sigma$ (GeV) 
and width $\Gamma_\sigma$ (GeV) of $\sigma$ meson, 
taken from Refs.~\cite{Mexico,BES,E791,PDG}
by following the previous Ref.~\cite{GKS}.} 
\label{kap_T_sigma}
\end{center}
\end{figure}

The momentum distribution for different components of thermal widths
and mean free paths are presented in the upper and lower panels of 
Fig.~(\ref{G_pi_k}). Upper panel shows that the pion
thermal width for $\pi\sigma$ (dotted line) and $\pi\rho$ (dashed line) 
loops become significant in the low and high momentum regions respectively. 
The corresponding mean free paths in the lower panel of the figure
expose this complementary feature of two resonances in more prominent way.
Here we notice that mean free path contribution from $\pi\sigma$ loop
is tending to diverge after $\vk\approx 0.1$ GeV while divergent nature
for $\pi\rho$ loop contribution is starting below the $\vk\approx 0.2$ GeV.
A phenomenological lesson from this fact is that
the low momentum pion will get the relevant dissipation by scattering
with $\sigma$ resonance whereas the dissipation of high momentum
pion will be coming from the $\pi\rho$ scattering in the medium.

Using the thermal width $\Gamma_{\pi(\pi M)}(\vk,T)$ in Eq.~(\ref{kap_pi}),
one can estimate the thermal conductivity, which possesses the temperature
dependence from not only the thermal width but also the Bose-enhanced phase
space factor of pions.
The contributions of $\pi\sigma$ (dotted line), $\pi\rho$ (dashed line)
loops and their total (solid line) in the thermal conductivity are
presented as functions of temperature in the Fig.~(\ref{kap_T_srT}),
where the upper and lower panels show the results with and without 
folding respectively.
The $\kappa_\pi(T)$ due to $\pi\rho$ loop
appears as a decreasing function and diverges in low temperature
region, where $\pi\sigma$ loop has a finite contributions. Because
of this compensating nature of two loops, $\kappa_\pi(T)$ for the mesonic
system is revealed as non-divergent well behaved function in the entire
temperature domain. Though the total thermal conductivity without
folding effect (lower panel) is exhibiting a divergent nature at very low temperatures 
($T<0.020$ GeV) but this divergence is cured completely after incorporating
the folding effect (upper panel). 

Similar to folding effect,
our results can also be influenced by cut-off parameter
$\Lambda$ of the hadronic form factor, which is taken in the $\pi\pi M$
vertex.
The results without form factor (dashed line) yield the minimum 
values of $\kappa$
vs $T$ curve, whereas results with form factor (solid line) exhibit
the enhanced values of $\kappa$ as shown in Fig.~(\ref{kap_T_comp}).
Owing to this fact, the thermal conductivity in our effective
hadronic model may vary within a band;
e.g. $\kappa=0.0086-0.013$ GeV$^2$ at $T=0.15$ GeV.

Besides this, 
our results may suffer from phenomenological 
uncertainty for fixing the parameters of $\sigma$ resonance
as discussed in our earlier work on shear viscosity~\cite{GKS}.
Longstanding controversies about the properties of $\sigma$ 
resonance seem to be settling to a consensus~\cite{Pelaez}
as the band of its mass-width values have been shortened from
$(m_\sigma=0.4-1.2$ GeV, $\Gamma_\sigma=0.6-1$ GeV)~\cite{PDG_old}
to $(m_\sigma=0.4-0.55$ GeV, $\Gamma_\sigma=0.4-0.7$ GeV)~\cite{PDG}.
Using the minimum (dash-dotted line) and maximum (dotted) values of
($m_\sigma,\Gamma_\sigma$) from latest PDG~\cite{PDG}, the band of
$\kappa$ without (lower panel) and with (upper panel) folding effect
are shown in Fig.~\ref{kap_T_sigma}.
The results of $\kappa$, for the mass-width parameters taken from BES~\cite{BES,BES2} (solid line)
and E791~\cite{E791} (dashed line) experiments, are also included in the 
Fig.~(\ref{kap_T_sigma}). 
Among them, the set of ($m_\sigma,\Gamma_\sigma$) from BES experiment 
is arbitrarily chosen
to generate all the previous results, shown in Fig.~(\ref{self_pi_M}) 
to (\ref{kap_T_comp}).
The list of effective coupling ($g_\sigma$) is provided in the Table
of Ref.~\cite{GKS}. 
From the upper panel of Fig.~(\ref{kap_T_sigma}),
$\kappa$ at $T=0.15$ GeV may vary from $0.012$ GeV$^2$ to $0.016$ GeV$^2$.
Hence, adding the uncertainty of $\sigma$ meson parameters with the form
factor effect, we get $\kappa(T=0.15$ GeV$)=0.0086-0.16$ GeV$^2$, which
is exhibiting quite large band of prediction of thermal conductivity
from the effective hadronic model. From the Table~(\ref{tab}), one can 
notice that the numerical values of $\kappa$ of earlier 
Refs.~\cite{Gavin,Greiner,Sarkar_kap,Nicola,Prakash,Davesne,Nam,Juan}
are within the band of uncertainty, contained in our results. 


\section{Summary}
\label{sec:concl}
In Summary, an estimation of thermal conductivity for pionic medium
has been done with the help of the effective hadronic model. We have
started with the standard expression of thermal conductivity 
in RTA approach, where inverse of pion relaxation time ($\tau_\pi$)
or pion thermal width $\Gamma_\pi$ has been evaluated from the imaginary
part of pion self-energy. Keeping in mind about the appearance of
$\sigma$, $\rho$ resonances in the $\pi\pi$ scattering cross section,
we have evaluated pion self-energy for $\pi\sigma$ and $\pi\rho$ loops
in the real-time formalism of thermal field theory. 
We get a non-monotonic momentum distribution of pion thermal width, which
has to be integrated out by the Bose-enhanced phase factor during
estimation of thermal conductivity. Our estimation has a large
band of uncertainty because of the various phenomenological parameters 
such as mass-width parameter of sigma meson, hadronic form factor etc.
Some~\cite{Gavin,Greiner,Sarkar_kap,Nicola,Prakash,Davesne,Nam,Juan} of
the earlier estimation of thermal conductivity for strongly interacting matter
in hadronic temperature domain are within our band of estimation.

%
{\bf Acknowledgment :}
Work is financed by Funda\c{c}\~ao de Amparo \`a Pesquisa do Estado de 
S\~ao Paulo - FAPESP, Grant Nos. 2012/16766-0.
I am very grateful to Prof. Gastao Krein for his academic
and non-academic support during my postdoctoral period in Brazil.
%
%
%

\end{document}